# Low-lying di-hadronic states in relativistic harmonic model

Ajay Kumar Rai[1,2]*, J N Pandya[2] and P C Vinodkumar[1]

[1]Department of Physics, Sardar Patel University, Vallabh Vidyanagar , Anand – 388 120, Gujarat, India.
[2]Applied Physics Department, Faculty of Technology and Engineering, M S University of Baroda,
Vadodara – 390 001, Gujarat, India.

E-mail: raiajayk@rediffmail.com

**Abstract:** Di-hadronic molecules such as di-meson, meson-baryon and di-baryon states are studied in the relativistic confinement model. We have computed the binding energy of the di-hadronic systems like *K-N, π-N, ρ-N etc.*, as penta-quark states, *2π, ρ-ρ, K-K*, K*-K*, D-D** etc. as tetra-quark states and *N-N, N-Δ, Δ-Δ* etc. as di-baryonic states using a molecular interaction provided by asymptotic expression of the confined gluon exchange potential. We find the lowest penta-quark state lies in the energy range of 1.180 - 2.247 GeV as predicted by other theoretical models. The exotic states such as $f_0$ (0.982), $h_1$ (1.170), $f_0$ (1.70), $f_2$ (1.565), $f_2$ (1.95), $\psi$(4.040) etc. are *identified* as the di-mesonic hadron molecules. The low-lying di-baryon molecular states found to be in the range of 1.990-2.907 GeV with their binding energies lying between 112-120 MeV.

## 1. Introduction

Exotic states of multi-quarks and gluons have been predicted by the theory of quantum chromodynamics and related phenomenological models [1-6]. The exotic states include those which do not fit in the *qq* and *qqq* states. So they include the four quark states, penta-quark states, hybrids such as *qqg* and *gg* or *ggg* glueball states. Though there are many candidates for the di-mesonic states, glueballs and hybrids [7], the recent discovery of a penta-quark state in the energy range of $1.54 \pm 0.01$ [8] GeV/c$^2$ has opened up renewed interest in the field of molecular hadrons in hadron physics. This penta-quark state is understood as a molecular meson-baryon resonance containing $uudd\bar{s}$ quarks [8-10].

The multi-quark system has previously been studied in the framework of Bag model [11] and nonrelativistic potential models [2, 12]. But there exist little success towards understanding the tetra-quark and penta-quark states as di-hadron molecules due to nonperturbative nature of QCD at the hadronic scale. It is widely believed that the QCD sum rule and the instanton based models are capable nonperturbative methods to extract the properties of hadrons [13, 14], but there exist little works on multiquark states, especially their systematic calculations. The time-delay method of meson-meson and meson-baryon scattering, provides useful information about spectroscopy of meson and baryon resonances [15, 16].

*Corresponding Author

Experimentally, there are several multi-quark candidates such as $f_1$ (1.42), $f_0$ (1.70), $f_2$ (1.565), $f_2$(1.96), $\psi$(4.040), X(3.87), $D^+_{sj}$ (2.632), Z+(1.54) *etc.*,[7, 8] reported.

Among the many theoretical attempts and approaches to explain the hadron properties based on its quark structure, very few were successful in predicting the hadronic properties starting from low flavour to heavy flavours. Apart from some recent algebraic attempts which lack the probable link with the theoretical approximations of qcd for the confinement of quarks [17] most of the properties of hadrons are reported in the potential framework. The nonrelativistic potential models with linear plus coulomb, power potential [18-20] etc. were successful at the heavy flavour sectors, while the nonrelativistic harmonic oscillator potentials [21] as well as the Bethe-Salpeter approach under harmonic confinement [22] were successful at the low flavour sectors.

Though the models based on instantons are derivable from first principles of QCD at the non-perturbative regime, they have been employed largely at the low flavour sector [14]. Similar to the instantons which are the fluctuations of the gluon field, that explains non-perturbative effects and they are working quite effectively in reproducing many features of the strong interactions, we have been able to provide a physical basis for the heuristic nature of the phenomenological confinement schemes with Lorentz scalar plus Lorentz vector character such as the relativistic harmonic confinement model (RHM) for quarks and the current confinement model (CCM) for gluons, based on an effective theory of small fluctuations around a background classical solution of Young-Mills theory [23] in the form of condensate of the background colour fields. These models have already been employed for the study of the magnetic moments of baryons, light flavour low-lying masses of hadrons and for the study of glueballs [24, 25]. They were also instrumental for the derivation of the residual effective (confined) one gluon exchange potential (COGEP) through Breit formalism under the Born approximation [26]. The RHM model with COGEP has also successfully employed for the study of nucleon-nucleon scattering phase shifts in quarks cluster calculations [27, 28] as well as in the study of masses and decay constants of mesons from low flavour to heavy flavours [29]. Recently, RHM has also been employed along with OGEP and III (Instanton Induced Interaction) for the study of the light flavour mesons [30].

We employ the relativistic harmonic model (RHM) which found to be one of the successful model for the study of various properties of hadrons and for the *N - N* interaction with the effective confined one gluon exchange interaction [27-29]. For the di - hadronic state, we consider a large *r* ($r \to \infty$) limit of the confined gluon propagator [29] that appear in the COGEP to get the weak residual confined gluon interaction similar to the Van-der Waals interaction of the atomic molecules. However, the strength of the residual confined gluon exchange interaction between the quarks within the hadron and between the hadrons of molecular state may be different. So, with the known $d\bar{s}uud$ penta-quark state as the *K-N* molecular hadron, we fix the molecular strength and compute all other low-lying di-meson tetra-quark states, meson-baryon penta-quark states and di-baryon hexa-quark states.

As we are employing the RHM scheme for the conventional hadronic mass calculations, the essential features of RHM is reviewed in Section 2 of this paper.

## 2. Model for mesons and baryons

In the relativistic harmonic confinement scheme (RHM) extended for heavy flavours (ERHM) [29], the quarks and antiquarks inside a hadron are under the action of a Lorentz vector plus scalar mean field potential of the form [24]

$$V(r) = \frac{1}{2}(1+\gamma_o) A^2 r^2 \quad (1)$$

Here, *A* is the confinement mean field parameter, $\gamma_o$ is the Dirac matrix and *r* is the mean distance of the quark from the hadronic center. The confined single particle state of the quarks under this mean field potential is described by the Dirac equation

$$[i\gamma^\mu \partial_\mu - m_q - V(r)]\psi_q(r) = 0, \quad (2)$$

where, the quark wave function $\psi_q(r)$ expressed in bispinor form $(\chi, \varphi)$ satisfy the coupled equations

$$(E - m_q - A^2 r^2)\chi = -i\sigma.\nabla\phi \quad (3)$$

and

$$(E + m_q)\phi = -i\sigma.\nabla\chi \quad (4)$$

Here, $m_q$ is the quark mass parameter and *E* is the energy eigenvalue of the single particle Dirac equation. So the two component form of the Dirac wave function can be written as

$$\psi_q(r) = N_q \begin{pmatrix} \chi(r) \\ -\dfrac{i\vec{\sigma}\cdot\vec{\chi}}{E+M_q} \end{pmatrix} \quad (5)$$

for quarks and

$$\psi_{\bar{q}}(r) = N_q \begin{pmatrix} -\dfrac{i\vec{\sigma}\cdot\vec{\chi}}{E+M_q} \\ -i\chi(r) \end{pmatrix} \quad (6)$$

for antiquarks. The equation satisfied by the upper component $\chi$ of $\psi_q(r)$ is now given by

$$[-\nabla^2 + (E+m_q)A^2 r^2]\chi = (E^2 + m_q^2)\chi \quad (7)$$

It is similar to the three - dimensional spherically symmetric harmonic oscillator equation with eigenvalue $E^2 - M_q^2$, and an energy dependent angular frequency (size parameter) $\Omega_N$, given by

$$\Omega_N = A(E_N + m_q)^{1/2} \quad (8)$$

Thus, we have the energy eigenvalue

$$E^2_N = \pm(m_q^2 + (2N+3)\Omega_N(q)) \quad (9)$$

where, N = 0, 1, 2, 3…

and the radial solution of eq. (7) is given by

$$R_{n\ell}(r) = \left\{ \begin{array}{l} \left[\dfrac{\Omega_N^{3/2}}{2\pi} \dfrac{n!}{\Gamma(n+\ell+3/2)}\right]^{1/2} \times \\ (\Omega_N^{1/2} r)^\ell \exp\left(\dfrac{-\Omega_N r^2}{2}\right) L_n^{\ell+1/2}(\Omega_N r^2) \end{array} \right\} \quad (10)$$

Here, the oscillator eigenvalue *2N + 3* occurs in the expression for $E_N^2$ instead of $E_N$ of a nonrelativistic oscillator potential. The radial quantum number *n* is related to $N$ by $N = 2n + l$. Using eq. (8) in eq. (9), we can get the single particle energy eigenvalue $E_N$ satisfying the equation

$$E_N^3 - M_q E_N^2 - M_q^2 E_N + M_q^3 - (2N+3)^2 A^2 = 0 \quad (11)$$

We can have an exact nonrelativistic reduction by eliminating the lower component $\phi$ from $\psi_q(r)$ totally through a transformation given by [27]

$$U\psi_q = \left\{ \begin{pmatrix} \dfrac{1}{1 + \dfrac{p^2}{(E+M_q)^2}} \end{pmatrix} \begin{bmatrix} 1 & \dfrac{\vec{\sigma}\cdot\vec{p}}{|E+M_q|} \\ -\dfrac{\vec{\sigma}\cdot\vec{p}}{|E+M_q|} & 1 \end{bmatrix} \right\} \begin{pmatrix} \chi \\ \phi \end{pmatrix} = \begin{pmatrix} \chi \\ 0 \end{pmatrix}$$

(12)

such that $\langle \psi_q | \psi_q \rangle = \langle \chi | \chi \rangle$ and are normalized. With such a reduction to nonrelativistic case, the single particle energy described by eq. (9) is identical for both quarks and antiquarks. Further in the limit of momentum $p \to 0$, the eigenvalue becomes proportional to the distance $r$ which is consistent with the lattice result of the linear behaviour of confinement potential. The single particle energy of the quark and antiquarks in this scheme, has been used to construct the masses of the mesons by taking into account of the spurious motion of the center of mass.

We adopt a scheme similar to the one which was used for the predictions of the low-lying baryonic masses by keeping the center of mass at the lowest eigenstate [24]. Then the intrinsic energies of the participating quarks are obtained by subtracting their contribution to the center of mass from their single particle energy. Accordingly, intrinsic energy of quarks in a $p$-quark system is given by [29]

$$\varepsilon_N(q)_{conf} = \sqrt{\left\{(2N+3)\Omega_N(q) + m_q^2 - \dfrac{3m_q}{\sum\limits_{i=1}^{p} m_q(i)}\Omega_0(q)\right\}} \quad (13)$$

The confined energies of the hadrons are then obtained by adding the respective intrinsic energies of the chosen flavour combinations. It is known that the residual Coulomb as well as the spin-hyperfine one gluon exchange interactions are also important in the predictions of hadronic masses containing heavy flavours [29]. Results of these computations are employed for the present study of di-hadronic systems.

Table 1: The model parameters, hadron masses and their size parameters from ERHM

| Mesons | Masses of mesons MeV | $\Omega_h$ MeV$^2$ | Baryons | Masses of baryons MeV | $\Omega_h$ MeV$^2$ |
|---|---|---|---|---|---|
| $\pi \& \rho$ | 149 & 767 | 63601 | $N \& \Delta$ | 939 & 1232 | 190802 |
| $K \& K^*$ | 507 & 921 | 77441 | $\Sigma \& \Sigma^*$ | 1181 & 1393 | 218482 |
| $D \& D^*$ | 1758 & 1964 | 227853 | – | – | – |

Confinement mean field parameters:

$A = 2166$ (MeV)$^{3/2}$
$M_u = M_d = 82.8$ MeV; $M_s = 357.5 MeV$; $M_c = 1428 MeV$;
$M_b = 4637 MeV$; CCM Parameter, $C = 50 MeV$ [29].

### 3. Binding energies of di-hadrons

The low-lying di-hadronic molecular mass consisting of either of di-meson tetra-quark states, meson-baryon penta-quark states and baryon-baryon hexa-quark states is given by

$$M^{h_1-h_2} = M_{h_1} + M_{h_2} + E_{BE}^{h_1-h_2}, \quad (14)$$

where $M_{h_1}$ and $M_{h_2}$ are the masses of the two hadrons, $h_1$ and $h_2$, $E_{BE}^{h_1-h_2}$, the molecular binding energy for the di-hadronic ($h_1$-$h_2$) system is obtained as

$$E_{BE}^{h_1-h_2} = \langle \psi_{h_1 h_2}(r_{12}) | V(r_{12}) | \psi_{h_1 h_2}(r_{12}) \rangle, \quad (15)$$

where $r_{12}$ is the relative coordinate of two hadrons. $V(r_{12})$ is the di-hadronic molecular potential. By considering residual interaction of the confined gluon similar to the Van-der Waals like interaction, and is assumed as that due to asymptotic expression ($r \to \infty$) of the residual confined one gluon exchange interaction with a strength ($k_{mol}$), we express it as [26, 31]

$$V(r_{12}) = -\dfrac{k_{mol}}{r_{12}} e^{-C^2 r_{12}^2 / 2}, \quad (16)$$

where $k_{mol}$ is the residual strength of the strong interaction molecular coupling and $C$ is the effective colour screening of the confined gluons.

The radial wave function for the ground state of the di-hadronic molecule is assumed as

$$\psi_{h_1 h_2}(r_{12}) = \left(4 \frac{\Omega_{12}^{3/2}}{\sqrt{\pi}}\right)^{1/2} e^{-\Omega_{12} r_{12}^2}, \quad (17)$$

where $\Omega_{12}$ is the size parameter of the di-hadronic state, and here we employ an additivity rule for the size parameter $\Omega_{12}$ as ($\Omega_1 + \Omega_2 = \Omega_{12}$), where $\Omega_1$ and $\Omega_2$ are the size parameter of the particular hadrons $h_1$ and $h_2$, respectively. Using eqs. (15-17) we get

$$E_B^{h_1-h_2} = \frac{4 k_{mol} \Omega_{12}^{3/2}}{\pi^{1/2}(\Omega_{12} + C^2/2)} \quad (18)$$

Here, in the case of meson-baryon system, $h_1$, $h_2$ correspond to the meson and baryon, respectively. With the known $\bar{s}uudd$ penta-quark state as the K-N molecular hadron [8], we fix the strength ($k_{mol} = 0.081$) of the residual confined gluonic interaction and predict all other low-lying di-hadronic states. The gluon confinement parameter C = 50 MeV has been chosen the same as that used earlier studies [29]. The size parameters $\Omega_1$, $\Omega_2$ and masses of the respective hadrons (mesons and baryons) employed here are from ERHM [29] and are listed in Table 1. The computed binding energy and masses for the system of di-mesons, meson-baryons and di-baryons systems with their binding energies are given in Tables 2, 3 and 4, respectively. The spin hyperfine interaction of the di-hadronic state is found to be negligibly small.

## 4. Conclusions

We have computed the binding energies and masses of exotic hadron resonances by treating them as di-hadronic molecules based on a confinement scheme for mesons and baryons. The present work is motivated by the recent experimental evidence for a narrow baryon resonance with strangeness $s = +1$, $J^P = (1/2)^-$, at the energy of $1.54 \pm 0.01$ GeV/c$^2$ that decays into $K^+$ and neutron [8]. This decay identifies the baryon resonance as a meson-baryon hadron molecule [8]. The penta-quark state as a single multiquark hadron can decays to many other possible channels which have not seen experimentally. This strongly supports the hadron molecular idea. Also such a multiquark state is expected to be at higher energy than observed at 1.5 GeV. However, the exact differences between them require more intense study. The present attempt on hadron molecules is just a first step in this direction.

The di-hadronic interaction has been taken as that due to the asymptotic expression of the confined one gluon exchange interaction among the quarks [28], while the hadronic masses have been computed using the ERHM model. The Van-der-Waals type of di-hadronic binding energies have been computed using the ERHM basis with a residual strong interaction strength $k_{mol}$. This parameter was fixed through the experimentally known penta-quark state as a K-N hadron molecule [8]. As we have considered the residual confine gluon exchange potential for the molecular interaction, it must be same for different compositions of the di-hadrons. Thus, the study provides us parameter-free predictions of various low-laying di-hadronic hadron molecular states. The computed results are shown in Tables 2, 3 and 4, respectively. Our predictions shown in Tables 2 and 3 are compared and identified with some of the experimentally known exotic hadronic states. These exotic states are those whose spin parity do not match with the expected quark- antiquark structure for mesons and 3-quark structure of baryons. Accordingly, the pseudoscalar di-mesonic and vector-vector di-mesonic combinations will have the parity and charge conjugation PC as ++ while for the combinations of pseudoscalar-vector di-meson state will have PC value +−, as shown in Table 2. The experimental low-lying candidates with the predicted $J^{pc}$ values are chosen for comparison.

Table 2: Low-lying masses and binding energies of di-meson molecule

| Di-hadron System $h_1$-$h_2$ | $J^{PC}$ | $E^{h1-h2}$ MeV | Di-mesons mass GeV | Expt. States [7] GeV |
|---|---|---|---|---|
| π-π | $0^{++}$ | 64.56 | 0.363 | $f_0$(0.4-1.2) |
| π-K | $0^{++}$ | 68.05 | 0.724 | - |
| π-ρ | $1^{+-}$ | 64.56 | 0.981 | - |
| K-K | $0^{++}$ | 71.36 | 1.085 | $f_0$(0.980) |
| π-K∗ | $1^{+-}$ | 68.05 | 1.138 | $h_1$(1.170) |
| ρ-K | $1^{+-}$ | 68.05 | 1.342 | $K_1$(1.273) |
| K-K∗ | $1^{+-}$ | 71.36 | 1.499 | $K_1$(1.400) |
| ρ-ρ | $0^{++}\ 1^{++}\ 2^{++}$ | 64.56 | 1.599 | $f_0$(1.590) |
| ρ-K∗ | $0^{++}\ 1^{++}\ 2^{++}$ | 68.05 | 1.756 | $f_0$(1.750) |
| K∗-K∗ | $0^{++}\ 1^{++}\ 2^{++}$ | 71.36 | 1.913 | $f_2$(1.950) |
| π-D | $0^{++}$ | 98.27 | 2.005 | – |
| π-D∗ | $1^{+-}$ | 98.27 | 2.211 | $D_1$(2.420) |
| K-D | $0^{++}$ | 100.59 | 2.366 | – |
| K-D∗ | $1^{+-}$ | 100.59 | 2.572 | $D_{sJ}$(2.573) |
| ρ-D | $1^{+-}$ | 98.27 | 2.623 | |
| K∗-D | $1^{+-}$ | 100.59 | 2.780 | – |
| ρ-D∗ | $0^{++}\ 1^{++}\ 2^{++}$ | 98.27 | 2.829 | – |
| K∗-D∗ | $0^{++}\ 1^{++}\ 2^{++}$ | 100.59 | 2.986 | – |
| D-D | $0^{++}$ | 123.06 | 3.639 | – |
| D-D∗ | $1^{+-}$ | 123.06 | 3.845 | $X$(3.870) |
| D∗-D∗ | $0^{++}\ 1^{++}\ 2^{++}$ | 123.06 | 4.051 | $\psi$(4.040) |

Table 3: Lowlying masses of penta-quarks states as meson-baryon negative parity baryon molecule

| Di-hadron System $h_1$-$h_2$ | $J^P$ | $E^{h_1-h_2}$ MeV | Penta-quark mass GeV | Other Theory. GeV |
|---|---|---|---|---|
| π-N | $(1/2)^-$ | 91.75 | 1.180 | 1.20 [8] |
| π-Σ | $(1/2)^-$ | 96.66 | 1.427 | Λ(1.405)[7] |
| π-Δ | $(3/2)^-$ | 94.75 | 1.473 | – |
| K-N | $(1/2)^-$ | 94.24 | 1.540 | $Z^+$(1.54)[8] |
| K-Σ | $(1/2)^-$ | 99.02 | 1.787 | Σ(1.750 [7] |
| ρ-N | $(1/2)^-,(3/2)^-$ | 91.75 | 1.798 | |
| K-Δ | $(3/2)^-$ | 93.24 | 1.833 | 1.85 [16] |
| K*-N | $(1/2)^-,(3/2)^-$ | 94.24 | 1.954 | – |
| ρ-Σ | $(1/2)^-,(3/2)^-$ | 96.66 | 2.045 | |
| ρ-Δ | $(1/2)^-,(3/2)^-,(5/2)^-$ | 91.75 | 2.091 | – |
| K*-Δ | $(1/2)^-,(3/2)^-,(5/2)^-$ | 94.24 | 2.247 | – |

Table 4: Low-lying masses of hexa - quarks states as di-baryon molecule

| Di-hadron system $h_1$-$h_2$ | $J^P$ | $E^{h_1-h_2}$ GeV | Hexa - quark masses GeV | Other theory GeV [32] |
|---|---|---|---|---|
| N-N | $0^+\ 1^+$ | 112.55 | 1.990 | - |
| N-Δ | $1^+\ 2^+$ | 112.55 | 2.284 | 2.17 |
| Δ-Δ | $0^+\ 1^+\ 2^+\ 3^+$ | 112.55 | 2.577 | 2.46 |
| N-Σ | $0^+\ 1^+$ | 116.59 | 2.237 | - |
| Δ-Σ | $1^+\ 2^+$ | 116.59 | 2.530 | - |
| Σ-Σ | $0^+\ 1^+$ | 120.49 | 2.482 | - |
| Σ-Σ* | $1^+\ 2^+$ | 120.49 | 2.695 | - |
| Σ*-Σ* | $0^+\ 1^+\ 2^+\ 3^+$ | 120.49 | 2.907 | - |

We found many $0^{++}$ di-mesonic states in the energy range of 0.363 GeV to 1.913 GeV in the light flavour ($u$, $d$, $s$) sector. Many of these states are identified with experimentally known exotic mesonic states. Accordingly, we identify $h_1$ (1.170 GeV) $1^{+-}$ and $f_0$(0.980 GeV) $0^{++}$ states as the π-K* and K-K di-mesonic states. Other di-mesonic states with charmed D-mesons are also being studied. The $D_1$(2.42 GeV), $D_{s1}$(2.540 GeV), X(3.870 GeV) and ψ(4.040 GeV) are identified here as π-D*, K-D*, D-D* and 2D* di-mesonic states respectively. Many of these predicted states could be experimentally seen.

The low-lying penta-quark states as the meson-baryon di-hadronic molecule naturally provide us negative parity baryons. These states in the u,d,s sector, have been studied and are tabulated in Table 3. The non-strange meson-baryon molecular resonances in the mass range of 1.180-2.247 GeV are predicted. The well known Λ(1.405 GeV) [9, 10] is very close to predicted π - Σ di-hadronic molecule. Its decay (100 %) in to Σ π mode [7] also suggest it to be the meson-baryon molecule. Other states that we could identify are the Σ(1.750 GeV) $(1/2)^-$ baryon or K - Σ(1.787 GeV) state. The resonance (1.850 GeV) observed by Kelker et al [16] in a time delay analysis of $K^+N$ reaction is close to our K - Δ di-hadronic molecular state . As Δ is just spin excitation of N, the time delay analysis of K-Δ can throw more information regarding some of these states.

The binding energies of the di-baryonic systems (NN to ΣΣ) in the u, d, s, sector tabulated in Table 4 are in the range of 112 - 120 MeV compared to 92 - 99 MeV of the meson-baryon penta-quark systems and 64 - 123 MeV of the di-mesonic tetra-quark systems . The predicted masses of the di-baryonic systems are in the range of 1.990 - 2.907 GeV as against other theoretical predictions of 2.17 - 2.46 GeV [32]. Many of these di-hadronic molecular states with the predicted spin-parity could be seen experimentally.

**Acknowledgment**
We express our thanks to Prof. S. C. Phatak (IOP, Bhubaneswar, India) and Prof. Atsushi Hosaka (RCNP, Osaka